\documentclass[aps,prd,reprint,amsmath,amssymb,twocolumn]{revtex4}
\usepackage{xcolor}
\usepackage{amsmath}
\usepackage{txfonts}
\usepackage{amssymb}
\usepackage{mathrsfs}
\usepackage{graphicx}
\usepackage{epsfig}
\usepackage{dcolumn}
\usepackage{bm}
\usepackage{hyperref}
\hypersetup{
    colorlinks=true,
    linkcolor=blue,
    urlcolor=blue,
    citecolor=blue,
    }
\begin{document}

\title{Could the stochastic gravitational wave background from newborn magnetars be detected by the advanced LIGO and Einstein Telescope?}

\author{Yu-Long Yan}
\affiliation{Institute of Astrophysics, Central China Normal University, Wuhan 430079, China
}
\author{Quan Cheng}%
 \email{qcheng@ccnu.edu.cn}
\affiliation{Institute of Astrophysics, Central China Normal University, Wuhan 430079, China
}%
\author{Xiao-Ping Zheng}%
\affiliation{Institute of Astrophysics, Central China Normal University, Wuhan 430079, China
}%
\author{Xiao-Yue Yu}%
\affiliation{Institute of Astrophysics, Central China Normal University, Wuhan 430079, China
}%

\date{\today}

\begin{abstract}

Newborn magnetars are important gravitational wave sources due to their ultra-strong magnetic fields and fast spins, and the entire population in the Universe may significantly contribute to the stochastic gravitational wave background (SGWB). In this work, we investigate the SGWB from newborn magnetars and assess its detectability by the advanced LIGO (aLIGO) and Einstein Telescope (ET) based on three typical formation mechanisms of magnetars, i.e., the $\alpha-\Omega$ dynamo, convective dynamo, and magnetic flux conservation. For the two dynamo scenarios, when calculating the SGWB, we creatively incorporate the anti-correlations between the magnetic fields and initial spin periods $P_{\rm i}$ with the initial dipole-field distribution of newborn magnetars. For the flux-conservation scenario, a bimodal lognormal form is adopted to describe the distribution of initial dipole fields, and all magnetars are assumed to have the same $P_{\rm i}$. Our results show that the SGWB from newborn magnetars may be undetectable by the aLIGO and ET if the magnetars are formed due to these mechanisms since the signal-to-noise ratio of the SGWB with respect to the ET for an observation time of one year is only 0.37 for the $\alpha-\Omega$ dynamo, $3\times10^{-4}$ for the convective dynamo, and at most 0.21 for the flux conservation. 

\end{abstract}

\maketitle

\section{Introduction}\label{Sec:I}

The stochastic gravitational wave background (SGWB) arises from the superposition of gravitational waves (GWs) emitted by numerous sources across the Universe. These GWs vary in frequency and intensity, ultimately forming a GW background that is analogous to the Cosmic Microwave Background \cite{2007gwte.book.....M}. The SGWB can primarily be classified into two categories: cosmological and astrophysical origins. The SGWB of cosmological origin is also dubbed as the primordial GW background, which is contributed by GW emissions from energy density fluctuations, cosmic phase transitions (e.g., the electroweak phase transition and quark-gluon plasma phase transition), cosmic strings, and inflation that occurred in the extremely early stages of the Universe \cite{1989NuPhB.319..747A,  2000csot.book.....V,2018PhRvD..97l3505C,2023PhRvL.131q1404G,2023JHEAp..39...81V,2024arXiv240104388R}. The SGWB of astrophysical origin is produced by the cumulative contributions of many individual astrophysical sources throughout the Universe, including mergers of compact binaries (e.g., binary neutron stars (NSs), binary black holes, and NS-black hole systems) \cite{2001MNRAS.324..797S,2006ApJ...642..455R,2012PhRvD..85j4024W,2013MNRAS.431..882Z,2016PhRvL.116f1102A,2017PhRvL.119p1101A,2021ApJ...915L...5A}, collapse and explosions of massive stars \cite{1999MNRAS.303..247F,2009MNRAS.398..293M,2022PhRvD.105f3022F}, and various instabilities of NSs (e.g., bar-mode, secular mode, f-mode, and r-mode instabilities) \cite{1995ApJ...442..259L,1998MNRAS.299.1059A,1999MNRAS.303..258F,2011ApJ...729...59Z,2012ApJ...761...63P,2022PhRvD.106f3007K}, as well as long-lived deformations of NSs caused by elastic stress in the crusts or internal magnetic fields \cite{2002PhRvD..66h4025C,2024arXiv240302066J,2006A&A...447....1R,2011MNRAS.411.2549M,2012PhRvD..86j4007R,2013PhRvD..87d2002W,2015MNRAS.454.2299C,2017PhRvD..95h3003C,2021Univ....7..381C}. Therefore, modeling and detection of the SGWB offer a unique avenue for probing the early Universe and the physics behind some astrophysical phenomena.

As a peculiar subclass of NSs, magnetars are considered to possess ultra-strong surface dipole magnetic fields whose typical strength is $\sim10^{14}$--$10^{15}$ G (see the \href{http://www.physics.mcgill.ca/~pulsar/magnetar/main.html}{McGill Online Magnetar Catalog} for details) \cite{2014ApJS..212....6O}. Furthermore, even stronger toroidal magnetic fields with strengths of $\sim2$--$100$ times the dipole fields may exist in the interior of magnetars (e.g., \cite{1992ApJ...392L...9D,2005ApJ...634L.165S,2009MNRAS.395.2162L,2009MNRAS.397..763B,2014PhRvL.112q1102M,2020SciA....6.2732R,2022A&A...668A..79B,2024EPJC...84.1043Y}). Results from axisymmetric magnetohydrodynamic (MHD) simulations showed that in the magnetar interior toroidal field may indeed be dominant although the internal stable field configuration possibly has a poloidal-toroidal twisted-torus shape \cite{2004Natur.431..819B,2006A&A...450.1097B,2009MNRAS.397..763B}. Therefore, if the magnetars have millisecond spin periods after birth \cite{1992ApJ...392L...9D,2001ApJ...552L..35Z}, the GW emission from their magnetic deformation could be remarkable \cite{2005ApJ...634L.165S,2009MNRAS.398.1869D,2024EPJC...84.1043Y}, making them promising targets for the advanced LIGO (aLIGO), advanced Virgo (aVirgo), KARGRA, and the third generation GW detectors, such as the Einstein Telescope (ET) and Cosmic Explorer \cite{2010CQGra..27s4002P,2017CQGra..34d4001A}. 

The origin of magnetars' strong magnetic fields is a long-standing puzzling issue and very worthy of discussion because different formation mechanisms of strong magnetic fields may result in different maximum strengths of the magnetic fields. An investigation into this issue can also help to figure out why magnetars are so different from other normal pulsars. Some theoretical work suggested that the magnetars' strong magnetic fields could be produced by the dynamo processes operate in nascent or newborn NSs (e.g., \cite{1992ApJ...392L...9D,1993ApJ...408..194T,2003ApJ...584..954A,2006Sci...312..719P,2014ApJ...786L..13C,2020SciA....6.2732R,2022A&A...668A..79B,2024arXiv241119328R}). Alternatively, magnetic flux conservation during the core collapse of highly magnetized massive progenitor stars (types O and B stars) that give birth to magnetars could also account for the formation of strong magnetic fields (e.g., \cite{2006MNRAS.367.1323F,2006MNRAS.370L..14V,2008MNRAS.389L..66F,2009MNRAS.396..878H}). In the dynamo scenarios, the eventual (saturation) strengths of a magnetar’s (both surface dipole and internal toroidal) magnetic fields after amplification are generally related to the magnetar’s initial spin period $P_{\rm i}$, and a smaller $P_{\rm i}$ will result in stronger fields since more stellar rotational energy could be extracted and converted into magnetic energy \cite{1992ApJ...392L...9D,2006Sci...312..719P,  2014ApJ...786L..13C,2020SciA....6.2732R,2022A&A...668A..79B}. However, in the magnetic flux-conservation scenario, the eventual strengths of the magnetar’s magnetic fields are mainly determined by the magnetization of its progenitor star, not essentially dependent on $P_{\rm i}$ of the magnetar. We clarify that the eventual dipole field after amplification is essentially the initial dipole field of the newborn magnetar, and so is the toroidal field. During the early evolution of the newborn magnetar, we neglect the decay of dipole and toroidal fields since their decay timescales are far beyond the magnetar's spin-down timescale \cite{1992ApJ...395..250G}. Hence, one should keep in mind that the dipole and toroidal fields of the newborn magnetar actually refer to their initial strengths.

In the context of toroidal-dominated internal fields, the amplitude of GWs from magnetic deformation of a single newborn magnetar is directly related to its toroidal field and spin frequency \cite{2009MNRAS.398.1869D,2017PhRvD..95h3003C,2024EPJC...84.1043Y}. Since the strengths of magnetic fields depend on $P_{\rm i}$ of the magnetar in the dynamo scenarios, the relation between the two may play a decisive role in estimating the single magnetar's GWs. Actually, when the dynamo processes terminate, the newborn magnetar population probably has different eventual strengths of dipole magnetic fields, which could be probed by pulsar population synthesis (e.g., \cite{2010MNRAS.401.2675P}). As a result, it is necessary to revisit the SGWB from newborn magnetars after involving (i) the relations between the strengths of magnetic fields and $P_{\rm i}$ of magnetars obtained based on various dynamo mechanisms, and (ii) the initial dipole-field distribution of newborn magnetars derived from pulsar population synthesis. In previous work, the SGWBs from newborn magnetars were roughly obtained assuming that all the magnetars have the same magnetic fields and $P_{\rm i}$ \cite{2011MNRAS.411.2549M,2013PhRvD..87d2002W,2015MNRAS.454.2299C,2017PhRvD..95h3003C}. Relatively exquisite results were also obtained after considering some specific forms for the distributions of $P_{\rm i}$ and magnetic fields of magnetars \cite{2012PhRvD..86j4007R}, however, neglecting the potential link between the former and the latter. Also, it is interesting to reinvestigate the SGWB from newborn magnetars after taking into account the fact that the magnetization of types O and B stars is distributed in a certain range \cite{2006MNRAS.367.1323F,2021MNRAS.504.5813M} if magnetars are indeed formed due to magnetic flux conservation. These may help to obtain a more realistic SGWB contributed by the entire newborn magnetars, and figure out whether this SGWB could be detected by the aLIGO and ET. 

The content of this work is organized as follows. Evolution of newborn magnetars, and the SGWB from magnetic deformation of these magnetars are introduced in Sec. \ref{Sec:II}. The initial dipole-field distributions of newborn magnetars in different formation scenarios are exhibited in Sec. \ref{Sec:III}. We also present the relations between the strengths of magnetic fields and $P_{\rm i}$ of magnetars obtained from two dynamo models in this section. Our results are shown in Sec. \ref{Sec:IV}. Finally, conclusions and discussions are given in Sec. \ref{Sec:V}.

\section{Evolution of newborn magnetars and their SGWB}\label{Sec:II}
\subsection{Evolution of newborn magnetars}
Before calculating the SGWB from magnetic deformation of newborn magnetars, it is necessary to investigate the evolution of a single newborn magnetar. Detailed calculations about this issue were performed in Refs. \cite{2009MNRAS.398.1869D,2015MNRAS.454.2299C,2018PhRvD..97j3012C}, and have recently been improved in \cite{2024EPJC...84.1043Y}. We follow the same method as that in \cite{2024EPJC...84.1043Y} to investigate the spin, tilt angle, and thermal evolutions of the newborn magnetar. As mentioned in Sec. \ref{Sec:I}, in the stellar interior even stronger toroidal field possibly exists, the newborn magnetar may thus be distorted into a prolate ellipsoid whose ellipticity is $\epsilon_{\rm B}=-1.6\times 10^{-4}(\bar{B_{\rm t}}/10^{16}~{\rm G})^2$ \cite{2002PhRvD..66h4025C}, where $\bar{B_{\rm t}}$ denotes the volume-averaged strength of the toroidal field. Following Ref. \cite{2024EPJC...84.1043Y}, the newborn magnetar is thought to lose angular momentum mainly through magnetic dipole (MD) radiation and GW emission from magnetic deformation. Other spin-down torques due to, for instance, r-mode and f-mode instabilities, and relativistic neutrino-driven wind are all neglected for simplicity (see \cite{2024EPJC...84.1043Y}). Considering that the newborn magnetar is probably embedded in a plasma-filled magnetosphere \cite{2006ApJ...648L..51S}, the evolution of its angular frequency $\omega$ can be expressed as \cite{2006ApJ...648L..51S,2000PhRvD..63b4002C,2024EPJC...84.1043Y}
\begin{eqnarray}
    \dot{\omega}=-\frac{B_{\rm d}^2R^6\omega^3}{6Ic^3}(1+\sin^2\chi)-\frac{2G\epsilon_{\rm B}^2I\omega^5}{5c^5}\sin^2\chi(1+15\sin^2\chi),\label{domega}
\end{eqnarray}
where $B_{\rm d}$, $R$, $\chi$, and $I=0.35MR^2$ are respectively the magnetar's surface dipole field, radius, magnetic tilt angle (the angle between the spin and magnetic axes), and moment of inertia with $M$ denoting its gravitational mass. In this work, we take typical values $M=1.4M_\odot$ and $R=12$ km.   

The newborn magnetar's GW emission is tightly related to its tilt angle $\chi$. Depending on the stellar temperature $T$, the evolution of $\chi$ could generally be divided into two stages \cite{2009MNRAS.398.1869D,2015MNRAS.454.2299C,2018PhRvD..97j3012C,2024EPJC...84.1043Y}. In the first stage, $T$ of the magnetar is so high that the stellar matter is in the liquid state and the neutrons in the core are non-superfluid. The free precession of the magnetar could be damped by the bulk viscosity of dense matter, leading to the increase of $\chi$ \cite{2009MNRAS.398.1869D}. As the newborn magnetar cools down, a solid crust will be formed on the surface, and the neutrons in the core will become superfluid \cite{2004ApJS..155..623P,2011PhRvL.106h1101P,2008LRR....11...10C}. If $\chi$ of the magnetar has not increased to $\pi/2$ in the first stage, the second evolutionary stage will begin, in which the internal viscosity from core-crust coupling could damp free precession of the magnetar, resulting in the increase of $\chi$. After involving the aligned torques from MD radiation and GW emission \cite{2000PhRvD..63b4002C,2009MNRAS.398.1869D,2014MNRAS.441.1879P}, the evolution of $\chi$ in both the first and the second stages can be given as \cite{2009MNRAS.398.1869D,2024EPJC...84.1043Y}

\begin{eqnarray}
    \begin{split}
    \dot{\chi} = &\frac{\cos \chi}{\tau_{\rm d} \sin \chi}- \frac{2G}{5c^5}I\epsilon_{\rm B}^2\omega^4\sin \chi \cos \chi(15\sin^2 \chi+1) \\ 
    &-\frac{B_{\rm d}^2R^6\omega^2}{6Ic^3}\sin \chi \cos \chi, \\
    \end{split}\label{chidot}
\end{eqnarray}
where $\tau_{\rm d}$ is the damping timescale of the magnetar's free precession caused by internal viscosities. In the first stage, we have $\tau_{\rm d}=\tau_{\rm bv}$ with $\tau_{\rm bv}$ representing the damping timescale of stellar free precession due to bulk viscosity. Following \cite{2009MNRAS.398.1869D}, the expression for $\tau_{\rm bv}$ is 
\begin{eqnarray}
    \tau_{\rm bv}\simeq 3.9\frac{\cot^2\chi}{1+3\cos^2\chi}\left(\frac{\bar{B_{\rm t}}}{10^{16}~{\rm G}}\right)^2\left(\frac{P}{1~{\rm ms}}\right)^2\left(\frac{T}{10^{10}~{\rm K}}\right)^{-6} ~{\rm s},\label{taubv}
\end{eqnarray}
where $P = 2\pi/\omega$ is the newborn magnetar's spin period. The damping timescale in the second stage is given by \cite{2024EPJC...84.1043Y}
\begin{eqnarray}
    \frac{1}{\tau_{\rm d}}=\frac{1}{\tau_{\rm bv}}+\frac{1}{\tau_{\rm cc}},\label{taud2}
\end{eqnarray}
where $\tau_{\rm cc}=\xi P/\left|\epsilon_{\rm B}\right|$ is the damping timescale of stellar free precession caused by core-crust coupling with $\xi$ denoting the number of precession cycles \cite{2005ApJ...634L.165S,1988ApJ...327..723A,2002PhRvD..66h4025C}. In the calculations, a typical value $\xi=10^4$ is used \cite{1988ApJ...327..723A,2002PhRvD..66h4025C,2024EPJC...84.1043Y}.   

For simplicity, the boundary between the first and second evolutionary stages is chosen to be $T=10^9$ K \cite{2024EPJC...84.1043Y}. Such a temperature is roughly equal to the one at which the newborn magnetar's solid crust is formed, and also the critical temperature for the onset of neutron superfluidity in the core \cite{2008LRR....11...10C,2018PhRvC..97a5804B,2018PhRvD..97j3012C,2024EPJC...84.1043Y}. The newborn magnetar cools down mainly through neutrino emission, its thermal evolution can be described by the formula below \cite{2024EPJC...84.1043Y} 
\begin{eqnarray}
    C_{\rm V}\frac{dT}{dt}=-L_{\nu,~\rm{MU}},\label{dTdt}
\end{eqnarray}
where $C_{\rm V} \approx 10^{39}(T/10^9~{\rm K})$ erg/K is the total specific heat of the magnetar, and $L_{\nu,~\rm{MU}} \approx 7\times 10^{39}(T/10^9~{\rm K})^8$ erg/s is the total luminosity of neutrino emission due to modified Urca processes \cite{2006NuPhA.777..497P}. 

The luminosity of GWs emitted by the newborn magnetar is \cite{2000PhRvD..63b4002C,2011MNRAS.411.2549M,2015MNRAS.454.2299C}
\begin{eqnarray}
    \frac{dE_{\rm gw}}{dt}=\frac{2G}{5c^5}\epsilon_{\rm B}^2 I^2\omega^6 \sin^2\chi(16\sin^2\chi+\cos^2\chi).\label{dEgwdt}
\end{eqnarray}
The first term on the rhs of Eq. (\ref{dEgwdt}) represents the luminosity of the GWs emitted at twice the spin frequency $\omega/\pi$ of the magnetar, while the second term denotes the luminosity of the GWs emitted at the spin frequency $\omega/2\pi$ of the magnetar. If the magnetar's tilt angle satisfies $0<\chi<\pi/2$, the two polarizations of the GWs from magnetic deformation contain components at both the spin frequency and twice the spin frequency of the magnetar. For details, one can refer to Equations (20) and (21), and Section 2.4 in \cite{1996A&A...312..675B}. In the special case of an orthogonal rotator ($\chi=\pi/2$), the GWs are emitted only at twice the spin frequency of the magnetar. Depending on the emitted GW frequency $\nu_{\rm e}$, the GW energy spectrum has the following form \cite{2015MNRAS.454.2299C}
\begin{eqnarray}
    \begin{split}
    \frac{dE_{\rm gw}}{d\nu_{\rm e}} &=\frac{4 \pi G}{5 c^{5}} \epsilon_{\rm B}^{2} I^{2} \omega^{6}\left|\dot{\omega}^{-1}\right| \sin ^{2}\chi\cos ^{2} \chi, \,\,\,\,\, \rm{for}\,\,\, \nu_{\rm e} \le \frac{\omega}{2\pi} \\
    &=\frac{32 \pi G}{5 c^{5}} \epsilon_{\rm B}^{2} I^{2} \omega^{6}\left|\dot{\omega}^{-1}\right| \sin ^{4} \chi, \,\,\,\,\, \rm{for}\,\,\, \frac{\omega}{2\pi}<\nu_{\rm e} \le \frac{\omega}{\pi}.
    \end{split}\label{dEgwdnu}
\end{eqnarray} 
 
\subsection{SGWB from newborn magnetars}
The SGWB is generally characterized by the dimensionless quantity $\Omega_{\rm GW}(\nu_{\rm obs})$, which denotes the distribution of dimensionless GW energy density versus the observed GW frequency $\nu_{\rm obs}$ by the detectors. To be specific, the SGWB produced by magnetic deformation of newborn magnetars generally can be expressed as \cite{2011MNRAS.411.2549M,2013PhRvD..87d2002W,2015MNRAS.454.2299C}
\begin{eqnarray}
    \begin{split}
    \Omega_{\mathrm{GW}}\left(\nu_{\mathrm{obs}}\right)=&\frac{8 \pi G \nu_{\mathrm{obs}}}{3 H_{0}^{3} c^{2}} \int_{0}^{z_{\mathrm{upp}}} \frac{R_{\rm s}(z)}{(1+z)E(\Omega, z)} \frac{d E_{\mathrm{gw}}}{d \nu_{\mathrm{e}}}d z,
    \end{split}\label{Omegagw}
\end{eqnarray}
where $\nu_{\rm obs}=\nu_{\rm e}/(1+z)$ with $z$ representing the cosmological redshifts of the sources. The upper limit of the redshift integration is determined by $z_{\rm upp}=\rm{min}(z_{\star},~\nu_{\rm e,max}/\nu_{\rm obs}-1)$ with $z_{\star}$ and $\nu_{\rm e,max}$ denoting the maximal redshift of the cosmic star formation rate (CSFR) model adopted and the maximal emitted frequency of the GWs, respectively. $R_{\rm s}(z)$ is the magnetar formation rate in the source frame, which can be expressed as
\begin{eqnarray}
R_{\rm s}(z)=\dot{\rho}_{*}(z)\int_{m_{\rm {min}}}^{m_{\rm {max}}}\Phi(m)dm \int_{B_{\rm d,1}}^{B_{\rm d,2}}P_{B_{\rm d}}(B_{\rm d};\mu,\sigma)dB_{\rm d}.\label{Rsz}
\end{eqnarray}
Following Ref. \cite{2006ApJ...651..142H}, the CSFR density $\dot{\rho}_{*}(z)$ in the above equation can be written as  
\begin{eqnarray}
   \dot{\rho}_{*}(z)=\frac{0.7(0.017+0.13 z)}{1+(z / 3.3)^{5.3}} \mathrm{M}_{\odot} \mathrm{yr}^{-1} \mathrm{Mpc}^{-3}.\label{rho}
\end{eqnarray}
The maximal redshift of this CSFR model is $z_{\star}=6$ \cite{2006ApJ...651..142H}. The Salpeter initial mass function in Eq. (\ref{Rsz}) has the form $\Phi(m)=Am^{-2.35}$ with $A$ denoting the normalization constant, which can be obtained by using $\int_{0.1M_\odot}^{125M_\odot}m\Phi(m)dm=1$. In this work, the standard $\Lambda$CDM cosmological model is used, thus we have $E(\Omega, z)=\sqrt{\Omega_{\rm m}(1+z)^3+\Omega_{\Lambda}}$ with $\Omega_{\rm m}=0.3$ and $\Omega_{\Lambda}=0.7$, and the Hubble constant $H_0=70~\rm{km}~\rm{s}^{-1}~\rm{Mpc}^{-1}$. The upper and lower limits on the mass of the progenitors that could produce NSs are chosen to be $m_{\rm min}=8M_\odot$ and $m_{\rm max}=40M_\odot$ \cite{2011MNRAS.411.2549M,2015MNRAS.454.2299C}. The term $\dot{\rho}_{*}(z)\int_{m_{\rm {min}}}^{m_{\rm {max}}}\Phi(m)dm$ in Eq. (\ref{Rsz}) thus represents the NS formation rate in the source frame. $P_{B_{\rm d}}(B_{\rm d};\mu,\sigma)$ in Eq. (\ref{Rsz}) is the dipole-field probability density function, which depicts the distribution of $B_{\rm d}$ of the newborn magnetar population. The specific forms of $P_{B_{\rm d}} (B_{\rm d};\mu,\sigma)$ in both the dynamo and the magnetic flux-conservation scenarios are presented in Sec. \ref{Sec:III}. $B_{\rm d,1}$ and $B_{\rm d,2}$ are respectively the lower and upper limits of $B_{\rm d}$ of newborn magnetars. The fraction of NSs that are born as magnetars therefore can be obtained via $\lambda_{\rm m}=\int_{B_{\rm d,1}}^{B_{\rm d,2}}P_{B_{\rm d}}(B_{\rm d};\mu,\sigma)dB_{\rm d}$.

In order to assess the detectability of the SGWB from newborn magnetars to the GW detectors, we need to estimate the signal-to-noise ratio (SNR). The optimized SNR of the SGWB for a given observation time $t_{\rm obs}$ can be expressed as
\begin{eqnarray}
    (\mathrm{S} / \mathrm{N})_{\mathrm{B}}=\left[\frac{9 H_{0}^{4} t_{\rm obs}}{50 \pi^{4}} \int_{0}^{\infty} \frac{\gamma^{2}\left(\nu_{\mathrm{obs}}\right) \Omega_{\mathrm{GW}}^{2}\left(\nu_{\mathrm{obs}}\right)}{\nu_{\mathrm{obs}}^{6} S_{h 1}\left(\nu_{\mathrm{obs}}\right) S_{h 2}\left(\nu_{\mathrm{obs}}\right)} d \nu_{\mathrm{obs}}\right]^{1/2},\label{snr}
\end{eqnarray}
where $\gamma(\nu_{\rm {obs}})$ is the normalized overlap reduction function, a quantity characterizing the sensitivity reduction to the SGWB caused by the separation and non-optimal orientations of the two detectors \cite{1993PhRvD..48.2389F}. Assuming that the two detectors are co-located and co-aligned, one has $\gamma(\nu_{\rm obs})=1$. $S_{h 1}(\nu_{\mathrm{obs}})$ and $S_{h 2}(\nu_{\mathrm{obs}})$ are respectively the power spectrum noise densities of the two detectors. In this work, we consider the cross-correlation of two identical detectors \cite{2008CQGra..25r4018R,2009LRR....12....2S,2011MNRAS.411.2549M}: 
(1) $S_{h1} = S_{h2} = S_{\mathrm{aLIGO}}$, corresponding to a network of two aLIGO detectors, and (2) $S_{h1} = S_{h2} = S_{\mathrm{ET}}$, corresponding to two ET detectors. In both cases, we assume an observation time of $t_{\rm obs}=1$ yr.

\section{Distributions of $B_{\rm d}$ and relations between field strengths and $P_{\rm i}$ of magnetars}\label{Sec:III}

By involving magneto-thermal evolution of NSs, Popov \textit{et al.} \cite{2010MNRAS.401.2675P} performed population synthesis simulations of various types of NSs, including thermally emitting isolated pulsars, normal radio pulsars, and magnetars. They proposed that the initial dipole fields of NSs generally follow a lognormal distribution, which can be written as     
\begin{eqnarray}
    P_B(B_{\rm d};\mu,\sigma_{\log B_{\rm d}})=\frac{1}{\sqrt{2\pi}\sigma_{\log B_{\rm d}}}\exp{\left\{-\frac{(\log B_{\rm d}-\mu)^2}{2\sigma_{\log B_{\rm d}}^2}\right\}}\label{distr1},
\end{eqnarray}
where $\mu=\langle \log(B_{\rm d}/{\rm G}) \rangle = 13.25$ is the best-fit central value, and $\sigma_{\log B_{\rm d}} = 0.6$ is the standard deviation \cite{2010MNRAS.401.2675P}. 

In dynamo scenarios, $B_{\rm d}$ and $P_{\rm i}$ of a magnetar are tightly related, and the relation between them depends on specific dynamo mechanisms considered. As a classical amplification mechanism of magnetic fields, the $\alpha-\Omega$ dynamo (hereafter model A) which arises due to intense neutrino-driven convection and differential rotation in a millisecond spinning protoneutron star can effectively amplify both its surface dipole and internal toroidal fields to magnetar-strength \cite{1992ApJ...392L...9D,1993ApJ...408..194T}. In this case, the newborn millisecond magnetar's toroidal field is approximately given as \cite{1992ApJ...392L...9D} 
\begin{eqnarray}
  \bar{B_{\rm t}}  \sim 3\times 10^{17}\left(\frac{P_{\rm i}}{1~\rm ms}\right)^{-1}~\rm{G}.
\end{eqnarray}\label{barBt}
The newborn magnetar's $B_{\rm d}$ is generally proportional to $\bar{B_{\rm t}}$. However, the ratio $\bar{B_{\rm t}}/B_{\rm d}$ is still unknown and may be within a rather wide range of $\sim2$--$100$ (e.g., \cite{1992ApJ...392L...9D,2005ApJ...634L.165S,2009MNRAS.395.2162L,2009MNRAS.397..763B,2014PhRvL.112q1102M,2020SciA....6.2732R,2022A&A...668A..79B,2024EPJC...84.1043Y}). In this work, we take $\bar{B_{\rm t}}/B_{\rm d}=100$. Such a ratio is possibly reasonable because the newborn magnetar with $P_{\rm i}=1$ ms could then have $B_{\rm d}\sim3\times10^{15}$ G, which is slightly larger than the strongest dipole field of magnetars measured to date \cite{2005ApJ...634L.165S}. Hence, in model A the upper limit of $B_{\rm d}$ in Eq. (\ref{Rsz}) is taken to be $B_{\rm d,2}=3\times10^{15}$ G. Without considering the effect of spin on the stability of the newborn magnetar's dipole field \cite{2006A&A...456..639G}, the lower limit is simply taken as $B_{\rm d,1}=10^{14}$ G, which generally represents the lower bound of typical dipole fields of magnetars \cite{2014ApJS..212....6O}. Thus for model A the toroidal fields are within $10^{16}\leq \bar{B_{\rm t}}\leq3\times10^{17}$ G. Combining Eq. (\ref{distr1}) with the values of $B_{\rm d,1}$ and $B_{\rm d,2}$ above, the proportion of magnetars to the total number of NSs is derived to be $\lambda_{\rm m}=10.55\%$.

The anti-correlations between the magnetic fields (both $B_{\rm d}$ and $\bar{B_{\rm t}}$) and initial spin ($P_{\rm i}$) of a protomagnetar have also been found when the convective dynamo (hereafter model B) plays a key role in amplifying the magnetic fields \cite{2020SciA....6.2732R}. The protomagnetar will contract and become a newborn magnetar as it cools down. Because of the contraction, the newborn magnetar's magnetic fields could generally be amplified by a factor of 4, while its spin period could be decreased by a factor of $6~{\rm ms}/2.3~{\rm ms}=2.6$ \footnote{The 6-ms period is the maximum spin period allowed for the protoneutron star with a radius of 20 km to achieve the strong-field dynamo branch. Such a protoneutron star will finally contract into a 
12-km newborn magnetar whose initial spin period is 2.3 ms. For details, one can refer to Fig. 4 in \cite{2020SciA....6.2732R} and related discussions.} with respect to that of the protomagnetar. Based on these scaling factors, by fitting Figure 4 in \cite{2020SciA....6.2732R} we can obtain the analytical formulas of $B_{\rm d}$ versus $P_{\rm i}$, and $\bar{B_{\rm t}}$ versus $P_{\rm i}$ for the newborn magnetar, which are respectively given as   
\begin{eqnarray}
    B_{\rm d}\simeq 4.60\times 10^{15}\left({P_{\rm i} \over 1~{\rm ms}}\right)^{-1.00}~{\rm G},
\end{eqnarray}

\begin{eqnarray}\label{barBtmb}
    \bar{B_{\rm t}}\simeq 2.81\times 10^{16}\left({P_{\rm i} \over 1~{\rm ms}}\right)^{-1.50}~{\rm G}.
\end{eqnarray}
Following Ref. \cite{2020SciA....6.2732R}, the initial spin periods of newborn magnetars are within $0.67\leq P_{\rm i}\leq10$ ms, in which the lower limit is determined by $P_{\rm c}/2.6=0.67$ ms with $P_{\rm c}=1.75$ ms representing the breakup spin periods of protomagnetars \cite{2020SciA....6.2732R}. From the allowed range of $P_{\rm i}$, we can obtain the lower and upper limits of $B_{\rm d}$ in model B as $B_{\rm d,1}=4.60\times10^{14}$ G and $B_{\rm d,2}=6.86\times10^{15}$ G. Combining Eq. (\ref{distr1}) and the interval of $B_{\rm d}$, the percentage of magnetars is derived to be $\lambda_{\rm m}=0.93\%$, an order of magnitude lower than that generally expected \cite{2021MNRAS.504.5813M}. To satisfy the expected percentage ($\sim10\%$) of magnetars \cite{2021MNRAS.504.5813M}, we artificially reduce the lower limit to $B_{\rm d,1}=10^{14}$ G (corresponds to $P_{\rm i}=46$ ms) without considering whether the convective dynamo could still work for such a slow initial spin. In this case, the percentage of magnetars with $10^{14}\leq B_{\rm d}\leq6.86\times10^{15}$ G is $10.56\%$.   

In the magnetic flux-conservation scenario, however, whether the magnetic fields of newborn NSs can be amplified to magnetar-strength predominantly depends on the magnetization of their progenitor stars, rather than their initial spins. Assuming simple flux conservation, the surface dipole field of the newborn NS is $B_{\rm d}=B_{\rm p}\left(R_\star/R\right)^2$, where $B_{\rm p}$ and $R_\star$ are respectively the magnetic field at the pole and the radius of the progenitor star. Makarenko \textit{et al.} \cite{2021MNRAS.504.5813M} found that the measured magnetic fields of both type O and type B stars follow a lognormal distribution though the central values and standard deviations are different. Based on the magnetic field distribution of O stars and the flux conservation hypothesis, the inferred mean dipole field of newborn NSs is $B_{\rm d}=5\times10^{14}$ ($1.7\times10^{12}$) G if the O stars are strongly (weakly) magnetized (model D in \cite{2021MNRAS.504.5813M}). It is worth noting that the magnetic fields of magnetic massive stars possibly follow a bimodal distribution, and $\sim10\%$ of all massive stars are strongly magnetized, the remaining $\sim90\%$ are weakly magnetized ones (see \cite{2021MNRAS.504.5813M} and references therein). Therefore, in principle magnetars and normal radio pulsars should respectively constitute $\sim10\%$ and $\sim90\%$ of the NS population \cite{2021MNRAS.504.5813M}. A bimodal lognormal form thus seems to be more aligned with the distribution of initial dipole fields of NSs, which can be written as \cite{2021MNRAS.504.5813M} 
\begin{eqnarray}\label{conser}
    \begin{split}
    P_B(B_{\rm d};\mu,\sigma)=&\frac{0.9}{\sqrt{2\pi}\sigma_1}\exp{\left\{-\frac{(\log B_{\rm d}-\mu_1)^2}{2\sigma_1^2}\right\}}\\
    &+\frac{0.1}{\sqrt{2\pi}\sigma_2}\exp{\left\{-\frac{(\log B_{\rm d}-\mu_2)^2}{2\sigma_2^2}\right\}},
    \end{split}
\end{eqnarray}
where the central values are $\mu_1=12.2$ and $\mu_2=14.7$, and the standard deviations are $\sigma_1=\sigma_2=0.6$ (see model D in \cite{2021MNRAS.504.5813M}). Using Eq. (\ref{conser}), the percentage of magnetars with $B_{\rm d,1}\leq B_{\rm d}\leq B_{\rm d,2}$ is derived to be $\lambda_{\rm m}=7.9\%$, where the lower and upper limits are respectively $B_{\rm d,1}=10^{14}$ G and $B_{\rm d,2}=3\times10^{15}$ G. 

It should be cautious that if the initial dipole fields of NSs follow the bimodal lognormal distribution above, the observed magnetar and normal pulsar populations cannot be simultaneously reproduced through NS population synthesis \cite{2021MNRAS.504.5813M}. As a result, Makarenko \textit{et al.} \cite{2021MNRAS.504.5813M} proposed that simple flux conservation cannot account for the origin of magnetic fields of NSs. Despite this disadvantage, we may still use Eq. (\ref{conser}) to estimate the initial dipole field distribution of magnetars since the percentage of magnetars is not overestimated for this distribution function. In the flux conservation scenario, little information is known for the toroidal fields of magnetars, though in principle, they possibly have comparable strengths to that of the dipole fields if the two components are comparable in the progenitor stars. This would naturally suppress GW emissions from newborn magnetars, resulting in a weak SGWB from newborn magnetars. To obtain a stronger SGWB from magnetars, in this scenario we assume stronger toroidal fields as $\bar{B_{\rm t}}=10B_{\rm d}$, and take lower strengths $\bar{B_{\rm t}}=5B_{\rm d}$ for comparison. Obviously, the ratios $\bar{B_{\rm t}}/B_{\rm d}$ are within the reasonable range of $\sim2$--$100$ (e.g., \cite{1992ApJ...392L...9D,2005ApJ...634L.165S,2009MNRAS.395.2162L,2009MNRAS.397..763B,2014PhRvL.112q1102M,2020SciA....6.2732R,2022A&A...668A..79B,2024EPJC...84.1043Y}). Finally, we assume that all magnetars have the same $P_{\rm i}$ for simplicity because the distribution of $P_{\rm i}$ of magnetars is uncertain, and $P_{\rm i}$ and $B_{\rm d}$ are probably not relevant in the flux conservation scenario. For the purpose of producing a strong enough SGWB, we take $P_{\rm i}=1$ ms for the magnetars. Such a fast spin may be unreasonable in this scenario. However, as we will see in Sec. \ref{Sec:IV}, even though $P_{\rm i}=1$ ms and $\bar{B_{\rm t}}=10B_{\rm d}$ are adopted, the resultant SGWB is still below the detection threshold of ET.  

\section{Results}\label{Sec:IV}

\begin{figure}[h]
    \centering
    \includegraphics[width=1\linewidth]{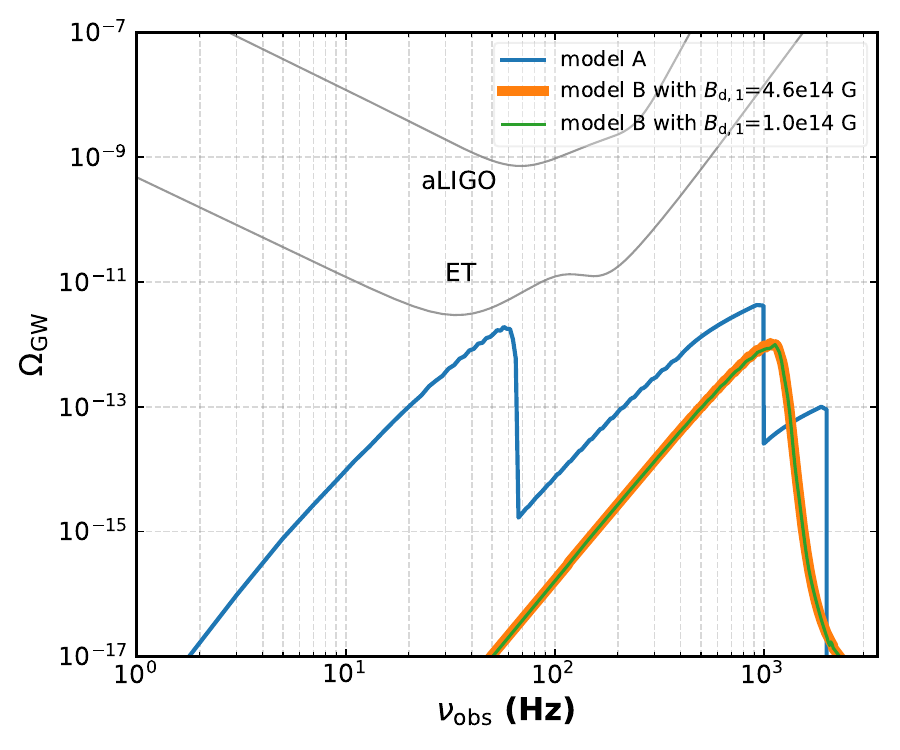}
    \caption{The dimensionless GW energy density $\Omega_{\rm GW}$ versus the observed frequency $\nu_{\rm obs}$ from newborn magnetars formed due to the two dynamo models. See the text for details.}
    \label{fig1}
\end{figure}

In Fig. \ref{fig1} we plot the dimensionless GW energy density $\Omega_{\rm GW}$ versus the observed frequency $\nu_{\rm obs}$ (the SGWB spectra) from newborn magnetars formed via the two dynamo models (models A and B as indicated in the legends) considered. In model B two different lower limits for the dipole fields of newborn magnetars $B_{\rm d,1}=10^{14}$ and $4.60\times10^{14}$ G are adopted. The detection thresholds of aLIGO and the planned ET calculated by using Eq. (136) in \cite{2009LRR....12....2S} and assuming an observation time of 1 yr are also presented (see the labels). 

The highest toroidal field in model A is $\bar{B_{\rm t}}=3\times10^{17}$ G if the newborn magnetar has an initial spin $P_{\rm i}=1$ ms. Such an initial period will result in a cut-off at the maximum GW frequency 2000 Hz in the background spectrum (see the blue curve) \cite{2006A&A...447....1R,2011MNRAS.411.2549M,2015MNRAS.454.2299C}. Following the same method used in \cite{2015MNRAS.454.2299C}, we can account for the sharp transitions at $1000$ and $\sim60$--$70$ Hz of the $\Omega_{\rm GW}$--$\nu_{\rm obs}$ curve in model A. As found in previous work \cite{2009MNRAS.398.1869D,2015MNRAS.454.2299C,2024EPJC...84.1043Y}, strong toroidal fields of $\bar{B_{\rm t}}\sim10^{17}$ G can significantly suppress the growth of $\chi$ of newborn magnetars in the first evolutionary stage, resulting in rather small tilt angles $\chi\simeq2^\circ$. The GW emission at twice the spin frequency of the magnetars is weakened owing to small $\chi$ [see Eq. (\ref{dEgwdnu})]. Thus the background emission at $1000\leq \nu_{\rm obs}\leq2000$ Hz is suppressed since it is only contributed by the emission at twice the spin frequency. The spectrum at $\nu_{\rm obs}\leq1000$ Hz, however, is contributed by emissions at both twice the spin frequency and the spin frequency of the magnetars when $\chi\neq\pi/2$. In the case of $\bar{B_{\rm t}}\sim10^{17}$ G, the early spin-down of the newborn magnetars is dominated by GW emission, by combining Eqs. (\ref{domega}) and (\ref{dEgwdnu}) we thus have 
\begin{eqnarray}
    \begin{split}
    \frac{dE_{\rm gw}}{d\nu_{\rm e}} &\simeq2\pi I\omega\cos ^{2} \chi, \,\,\,\,\, \rm{for}\,\,\, \nu_{\rm e}\leq1000~Hz \\
    &\simeq 16\pi I\omega\sin^2\chi, \,\,\,\,\, \rm{for}\,\,\, 1000\leq \nu_{\rm e}\leq2000~Hz.
    \end{split}\label{dEgwdnu2}
\end{eqnarray}
Obviously, the GW energy spectrum is enhanced by a factor of $\cot^2\chi/8\sim10^2$ at $\nu_{\rm e}=1000$ Hz for $\chi\simeq2^\circ$. We can thus observe a sharp transition at 1000 Hz in the background spectrum, as shown by the blue curve in Fig. \ref{fig1}. We also find that though newborn magnetars with $\bar{B_{\rm t}}\sim10^{17}$ G only account for a small proportion of the newborn magnetar population, the suppression in $\Omega_{\rm GW}$ at $1000\leq \nu_{\rm obs}\leq2000$ Hz as found in \cite{2015MNRAS.454.2299C} is still presented when the dipole field distribution is involved. 

The tilt angles of newborn magnetars with $\bar{B_{\rm t}}\sim10^{17}$ G in model A can increase to $\pi/2$ only when the core-crust coupling become effective in damping of free precession of these magnetars. At this point, they generally spin down to $\sim30$--$35$ Hz. When $\chi=\pi/2$ is achieved, GWs are emitted only at twice the spin frequency of the magnetars. Combining Eqs. (\ref{domega}) and (\ref{dEgwdnu}), and considering that the magnetars spin down mainly through MD radiation at this point, we have 
\begin{eqnarray}
    \begin{split}
    \frac{dE_{\rm gw}}{d\nu_{\rm e}} &\simeq\frac{96\pi G\epsilon_{\rm B}^2I^3\omega^3}{5c^2B_{\rm d}^2R^6}, \,\,\,\,\, \rm{for}\,\,\, \nu_{\rm e}\lesssim60-70~Hz \\
    &\simeq \frac{24\pi G\epsilon_{\rm B}^2I^3\omega^3\sin^2\chi}{5c^2B_{\rm d}^2R^6} , \,\,\,\,\, \rm{for}\,\,\, \nu_{\rm e}\gtrsim60-70~Hz.
    \end{split}\label{dEgwdnu3}
\end{eqnarray}
Therefore, the GW energy spectrum is enhanced by a factor of $4/\sin^2\chi\sim10^3$ at $\nu_{\rm e}\simeq60$--$70$ Hz for $\chi\simeq2^\circ$. This naturally accounts for the sharp transition in $\Omega_{\rm GW}$ at $\nu_{\rm obs}\sim60$--$70$ Hz in model A (see Fig. \ref{fig1}).

Although the sharp transitions at 1000 and $\sim60$--$70$ Hz in the SGWB spectrum of model A are an indication of the tilt angle evolution of newborn magnetars with $\bar{B_{\rm t}}\sim10^{17}$ G, the SGWB spectrum is well below the detection threshold of aLIGO, and also beneath that of the planned ET. To quantitatively estimate the detectability of the SGWB by the ET, we also calculate the SNR and show the result in Tab. \ref{tab:table1}. The corresponding SNR is 0.37, several times below the detection threshold 2.56 for the ET \cite{2011MNRAS.411.2549M,2017PhRvD..95h3003C}. The results suggest that the SGWB from newborn magnetars may be undetectable even using the ET if the magnetars' strong magnetic fields are produced due to the $\alpha-\Omega$ dynamo.

\begin{table}[h]
\caption{\label{tab:table1} The SNR of the SGWB with respect to the ET from newborn magnetars formed due to the two dynamo models (models A and B), and flux conservation. In the flux-conservation scenario, we respectively adopt $\bar{B_{\rm t}}=10B_{\rm d}$ and $\bar{B_{\rm t}}=5B_{\rm d}$ for the newborn magnetars. All SNRs are derived assuming an observation time of 1 yr.}
\begin{ruledtabular}
\begin{tabular}{lcr}
\multicolumn{3}{c}{SNR} \\[0.15cm]
\colrule
Dynamo & 0.37 (model A) & $3\times10^{-4}$ (model B) \\[0.15cm]
Flux Conservation & ~~0.21 ($\bar{B_{\rm t}}=10B_{\rm{d}}$) & $1.88\times10^{-2}$ ($\bar{B_{\rm t}}=5B_{\rm{d}}$)\\
\end{tabular}
\end{ruledtabular}
\end{table}

In contrast, the newborn magnetars in model B generally have weaker toroidal fields, which are within $8.89\times10^{14}\leq \bar{B_{\rm t}}\leq5.12\times10^{16}$ G or $9.00\times10^{13}\leq \bar{B_{\rm t}}\leq5.12\times10^{16}$ G, depending on the lower limit $B_{\rm d,1}$ adopted (see Eq. (\ref{barBtmb}) and the related discussions). The growth of $\chi$ in the first evolutionary stage of these newborn magnetars is very mildly or even totally not suppressed, thus $\chi=\pi/2$ is realized soon after the birth of the magnetars. The SGWB spectrum in model B does not exhibit the multi-peak feature as observed in model A (see Fig. \ref{fig1}). We can also find that varying $B_{\rm d,1}$ from $4.60\times10^{14}$ to $10^{14}$ G, the SGWB spectrum remains unchanged. Hereinafter, the SGWB spectrum in model B refers to the spectrum derived by using $B_{\rm d,1}=4.60\times10^{14}$ G. Compared to that in model A, the SGWB spectrum in model B is weaker in most frequency band, and the SNR of the latter with respect to the ET is only $3\times10^{-4}$ (see Tab. \ref{tab:table1}), suggesting that the SGWB from newborn magnetars could not be detected by the ET if the convective dynamo is responsible for the formation of strong magnetic fields of magnetars. In previous work, the SGWB from newborn magnetars were obtained by assuming that all the magnetars have the same $P_{\rm i}$ and magnetic fields \cite{2011MNRAS.411.2549M,2013PhRvD..87d2002W,2015MNRAS.454.2299C,2017PhRvD..95h3003C}. Compared to these results, the background spectra in models A and B are remarkably weaker.

\begin{figure}[h]
    \centering
    \includegraphics[width=1\linewidth]{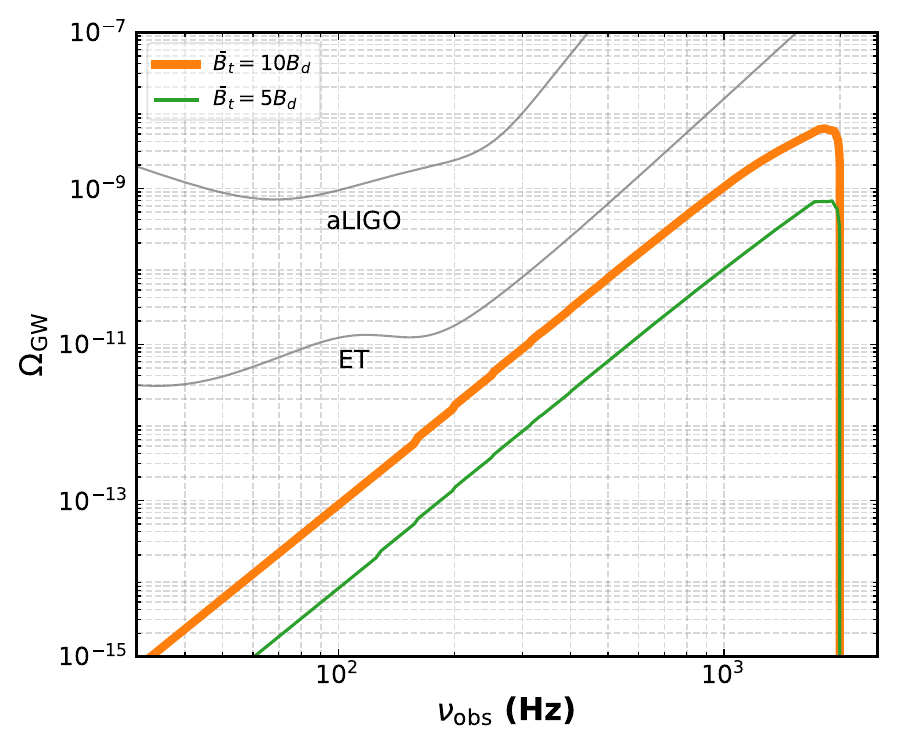}
    \caption{$\Omega_{\rm GW}$ versus $\nu_{\rm obs}$ from newborn magnetars formed due to magnetic flux conservation. The colored curves show the SGWB spectra obtained by assuming $\bar{B_{\rm t}}=10B_{\rm d}$ and $\bar{B_{\rm t}}=5B_{\rm d}$ for the newborn magnetars (see the legends).}
    \label{fig2}
\end{figure}

We show the SGWB from newborn magnetars in Fig. \ref{fig2}, assuming that the magnetars' strong magnetic fields are produced due to flux conservation. The background spectra are obtained using different scaling relations between $\bar{B_{\rm t}}$ and $B_{\rm d}$: $\bar{B_{\rm t}}=10B_{\rm d}$ and $\bar{B_{\rm t}}=5B_{\rm d}$ (see the legends) for the newborn magnetars. Since in this case the newborn magnetars have relatively low $\bar{B_{\rm t}}$, their $\chi$ can rapidly increase to $\pi/2$ in the first evolutionary stage, thus no multi-peak feature is observed in the resultant SGWB spectrum, similar to that in model B. Even all newborn magnetars have initial spin periods $P_{\rm i}=1$ ms and toroidal fields $\bar{B_{\rm t}}=10B_{\rm d}$, the SGWB spectrum lies well below the detection threshold of aLIGO, and also falls beneath that of the ET. The corresponding SNR of the spectrum with respect to the ET is 0.21 for $\bar{B_{\rm t}}=10B_{\rm d}$, while only $1.88\times10^{-2}$ for $\bar{B_{\rm t}}=5B_{\rm d}$, as listed in Tab. \ref{tab:table1}. Both of the SNRs are below the detection threshold 2.56 for the ET \cite{2011MNRAS.411.2549M,2017PhRvD..95h3003C}. Therefore, in the flux-conservation scenario, the SGWB from newborn magnetars also could not be detected by the aLIGO and ET. We note that the distributions of $P_{\rm i}$ and magnetic fields, however, without considering possible correlations between them, were involved in the calculations of the SGWB from magnetars in \cite{2006A&A...447....1R,2012PhRvD..86j4007R}. Their results slightly differ from the spectra obtained in the flux-conservation scenario because different formulas for the magnetic field distribution are adopted. In conclusion, if newborn magnetars are formed due to the $\alpha-\Omega$ dynamo, convective dynamo, or magnetic flux conservation, it is unlikely to detect the SGWB from these magnetars by the aLIGO, and even ET.

\section{Conclusions and discussions}\label{Sec:V} 

Whether the SGWB from newborn magnetars could be detected by the aLIGO and ET is still a topic under debate though previous results showed that it could be detectable by the ET (e.g., \cite{2011MNRAS.411.2549M,2013PhRvD..87d2002W,2015MNRAS.454.2299C,2017PhRvD..95h3003C}). In their calculations, all newborn magnetars are assumed to have the same $P_{\rm i}$ and magnetic fields, which is probably unrealistic. Actually, $P_{\rm i}$ and magnetic fields are generally correlated in view of the dynamo mechanisms that can account for the origin of strong magnetic fields of magnetars. In this work, we investigate the SGWB from newborn magnetars and estimate its detectability by the aLIGO and ET assuming that the magnetars are formed due to the $\alpha-\Omega$ dynamo, convective dynamo, or magnetic flux conservation. For the two dynamo scenarios, the anti-correlations between the magnetic fields and $P_{\rm i}$, together with the initial dipole-field distribution [Eq. (\ref{distr1})] of newborn magnetars are involved in our calculations. While for the flux-conservation scenario, we use the bimodal lognormal form [Eq. (\ref{conser})] to depict the distribution of initial dipole fields and assume that all magnetars have the same $P_{\rm i}$. To obtain the GW energy spectrum emitted by a single newborn magnetar, we study its tilt angle, spin, and thermal evolutions in detail based on the model in \cite{2024EPJC...84.1043Y}. Our results show that the SGWB from newborn magnetars could not be detected by the aLIGO and even ET if their strong magnetic fields indeed originate from the $\alpha-\Omega$ dynamo and convective dynamo. The SNRs of the background spectra with respect to the ET for an observation time of 1 yr are 0.37 and $3\times10^{-4}$ for the two dynamo scenarios, respectively. The SGWB in the flux-conservation scenario is also possibly undetectable by the aLIGO and ET because the SNR of the spectrum with respect to the ET for one-year observation time is only 0.21 even when all newborn magnetars have $\bar{B_{\rm t}}=10B_{\rm d}$ and $P_{\rm i}=1$ ms. Therefore, the detection prospect of the SGWB from newborn magnetars using the ET may not be promising. 

We note that before giving a definite conclusion on the detection prospect of the SGWB by the ET, more exquisite investigations are still needed. First, the effect of the initial dipole-field distribution of newborn magnetars on the SGWB deserves further study, especially when considering that the initial dipole-field distribution [Eq. (\ref{distr1})] was obtained without taking into account the GW emission from magnetic deformation of magnetars \cite{2010MNRAS.401.2675P}. Second, other mechanisms may also account for the origin of magnetars' strong magnetic fields \cite{2003ApJ...584..954A,2006Sci...312..719P,2014ApJ...786L..13C,2022A&A...668A..79B,2024arXiv241119328R}. It is thus necessary to investigate the dependence of magnetic fields on $P_{\rm i}$ of newborn magnetars in these scenarios, and what the SGWB would be if magnetars could be produced by these mechanisms. In fact, the strong magnetic fields of magnetars could be produced by a series of processes, such as magnetic flux conservation, the stationary accretion shock instability, $\alpha$–$\Omega$ dynamo, and convective dynamo \cite{2018PhRvD..97j3012C,2016MNRAS.460.3316R}. As a result, when calculating the SGWB from newborn magnetars, it is necessary to take into account at least two or more amplification mechanisms of magnetic fields, for instance, a combination of magnetic flux conservation and the $\alpha$–$\Omega$ dynamo. However, the relation between the magnetic fields and $P_{\rm i}$ of newborn magnetars in this case is probably more complicated and MHD simulations covering the core collapse of massive stars and the formation of newborn magnetars are required to address this issue, which is beyond the scope of this work. We expect a more robust result once the issues above could be resolved.

\begin{acknowledgments}
We gratefully thank the anonymous referee for helpful comments. This work is supported by the National Natural Science Foundation of China (Grant No. 12033001, No. 12473039, and No. 12003009), the National SKA program of China (Grant No. 2020SKA0120300), and the Fundamental Research Funds for the Central Universities (Grant No. 30106250142).
\end{acknowledgments}

\bibliographystyle{apsrev4-2}
\bibliography{reference}

\end{document}